# Assessing the stability fields of molecular and polymeric $CO_2$

Alexander F. Goncharov[1], Elena Bykova[1,2], Maxim Bykov[1,3], Eric Edmund[1], Jesse S. Smith[4], Stella Chariton[5], and Vitali B. Prakapenka[5]

[1] Earth and Planets Laboratory, Carnegie Institution for Science, Washington, DC 20015, USA

[2] Bayerisches Geoinstitut, University of Bayreuth, Universitätsstrasse 30, D-95447, Bayreuth, Germany

[3] Institute of Inorganic Chemistry, University of Cologne, Greinstrasse 6, 50939 Cologne, Germany

[4] HPCAT, X-ray Science Division, Argonne National Laboratory, Argonne, Illinois 60439, USA

[5] Center for Advanced Radiation Sources, The University of Chicago, Chicago, Illinois 60637, USA

**We investigated the stability of polymeric $CO_2$ over a wide range of pressures, temperatures, and chemical environments. We find that the $I\bar{4}2d$ polymeric structure, consisting of a three-dimensional network of corner sharing $CO_4$ tetrahedra, forms at 40-140 GPa and from a $CO-N_2$ mixture at 39 GPa. An exceptional stability field of 0 to 286 GPa and 100 to 2500 K is documented for this structure, making it a viable candidate for planetary interiors. The stability of the tetrahedral polymeric motif of $CO_2$-V is a consequence of the rigidity of $sp^3$ hybridized orbitals of carbon in a closed-packed oxygen sublattice.**

## I. INTRODUCTION

The high-pressure behavior of $CO_2$ molecules is fundamental to understanding deep planetary interiors as both the constituent elements are abundant in the Universe. $CO_2$ is an important component of planetary atmospheres such as Venus and Mars [1] and interiors such as the Earth's mantle [2-5]. The physical state and chemical transformations of $CO_2$ in planetary interiors under extreme pressure-temperature (P-T) conditions are not well understood especially toward ultrahigh P-T conditions, where possible physico-chemical transformations are not sufficiently explored.

The $CO_2$ molecule is linear at ambient conditions, owing to its electronic configuration where *sp*-hybridized electrons of the two C=O double bonds repel each other. This configuration makes $CO_2$ one of the most stable molecules at ambient conditions. However, at high pressures, the high-order bonds destabilize due to their increasing electron kinetic energy which becomes unfavorable compared to single-bonded configurations. This results in the transformation of molecular $CO_2$ to polymeric $CO_2$ with single-bonded carbon atoms which are tetrahedrally coordinated [6-8]. The transition from a molecular to a polymeric phase is hindered because of a substantial kinetic barrier and thus heating or other energy insertion is needed to accelerate this process [9-11]. The structure of the polymeric phase has been determined as tetragonal $I\bar{4}2d$ ($CO_2$-V) [7, 8, 12], consisting of a three-dimensional network of corner sharing $CO_4$ tetrahedra. Other structures of polymeric $CO_2$ have been proposed [6, 13, 14]. Additionally, it has been reported that six-fold coordinated $CO_2$ (like stishovite $SiO_2$) [15] and an "ionic" polymeric crystal form (i-$CO_2$) consisting of $CO_3$ and CO structural units [16], form at 50 GPa and 600 K and 85 GPa and 1700 K, respectively. In contrast, a recent diamond anvil cell (DAC) investigation up to 120 GPa combined with laser heating up to 6400 K showed only the presence of $CO_2$-V [17]. Theoretical calculations predict the stability of $I\bar{4}2d$ polymeric $CO_2$ up to at least 200 GPa, and at higher pressures a layered $P4_2/nmc$ phase is expected to become more stable [18, 19], where the fourfold carbon coordination persists.

At low pressures, polymeric $CO_2$-V was found to exist as a metastable phase upon decompression down to 4-10 GPa at 295 K and was reported to be quenchable to ambient pressure at 191 K [13]. This latter result was disputed in Ref. [20], where pressure measurements at low temperatures were criticized and a pressure of 1.9 GPa (instead of ambient) at 191 K was inferred for these measurements based on the reported Raman frequency of $CO_2$-I (dry ice), which is a stable phase at 0.3-10 GPa at 295 K. Recovery of polymeric $CO_2$ to ambient pressure even at low temperatures could, in principle, open a possibility for its technological applications [21], where $CO_2$ polymeric crystalline and amorphous [22] polymorphs could have superior properties compared to chemically similar $SiO_2$ quartz, for example, for creating stronger ultrafast transducers.

In the C-O system, molecular $CO_2$ represents a true thermodynamically stable compound at the bottom of the convex hull. Carbon monoxide CO (which also polymerizes at fairly low pressures) is the only other molecule which is close to the convex hull, however theoretical calculations show that this composition becomes less competitive at high pressures [23]. Nevertheless, other compounds in the C-O binary or within more complex mixtures may become more stable. For example, CO-$N_2$ mixtures are reported to form a CON$_2$ polymeric state above 45 GPa [24], which is synthesized at a lower pressure than for pure $N_2$. A theoretical structure search in the C-O-N system found a $Cmc2_1$-$C_2N_2O$ polymeric compound to be stable between 20 and 100 GPa [23, 25] alongside other metastable compounds with different compositions, which are stable in the CO-N composition space [23].

In this work, we explored the stability field of $CO_2$ over a wide P-T range of 1 Pa up to 286 GPa and 100-3000 K, where we find the $CO_2$-V structure to be the only (meta)stable polymeric phase. Dry ice $CO_2$-I coexists in the limits of low P-T conditions. These results suggest a possibility to recover polymeric $CO_2$-V to ambient pressure at low temperatures (cf. Ref. [20]) and demonstrate the exceptional stability of this phase up to 300 GPa and at least 2500 K suggesting a possibility that this material can exist in deep planetary interiors. Moreover, we find that a CO-$N_2$ mixture at 39 GPa reacts at high T above 2500 K yielding polymeric $CO_2$-V and diamond suggesting that the reported CON$_2$ polymeric state [24] at 50 GPa is actually a mixture of polymeric $CO_2$-V and common molecular ε-$N_2$. This finding demonstrates the high chemical stability of $CO_2$ and the challenges in creating ternary chemical compounds in CO-bearing systems.

## II. EXPERIMENTAL METHODS

The experiments were performed in diamond anvil cells equipped with Boehler-Almax and standard type diamond anvils with 200-300 µm culet sizes and toroidal anvils with 40 µm culet sizes for high (~40 GPa) and ultrahigh (up to 300 GPa) pressure ranges, respectively. Pre-indented rhenium foil was used as a gasket. Gold flakes were positioned in the sample cavity as laser absorbers to heat the sample. Laser heating was performed with a near infrared 1064-nm fiber laser using a double-side system with flat-top focusing at GSECARS (APS, ANL)[26]. $CO_2$ was loaded as a supercritical fluid at 0.01 GPa at room temperature in a gas loader. In one experiment CO was mixed with $N_2$ in an approximate ratio of 1:1 and gas loaded at 0.14 GPa. Polymeric $CO_2$ was synthesized in a laser heated DAC at 39 -140 GPa. The $I\bar{4}2d$ $CO_2$-V polymeric phase has been documented on all occasions as described below. XRD measurements were performed at GSECARS at high pressures up to 286 GPa and at HPCAT for low temperature studies (~100 K). Measurements at low temperatures were performed upon decompression using a double-diaphragm decompression attachment [27]. Concomitant Raman measurements were performed at the offline system at GSECARS [28] using 532 and 660 nm laser lines. Pressure was measured using the Equation of State (EOS) of gold (Au) determined in Refs. [29, 30] at high pressures up to 286 GPa and room temperatures and at pressures below 40 GPa and low temperatures, respectively. Measurements of the spectral position of the Raman signal of the stressed diamond [31] in concomitant XRD/Raman measurements yielded consistent pressure values.

For single crystal XRD measurements at HPCAT, we used monochromatic X- ray radiation with $\lambda = 0.3445$ Å and beam size of ~5 × 5 μm$^2$. Diffraction images were measured by a Pilatus 1 M pixel detector. For the single-crystal XRD measurements samples were rotated around a vertical ω-axis in a range of ±30° with an angular step Δω = 0.5° and an exposure time of 5 s/frame. For analysis of the single-crystal diffraction data we used the CrysAlisPro software package. To calibrate an instrumental model in the CrysAlisPro software, i.e., the sample-to-detector distance, detector's origin, offsets of goniometer angles, and rotation of both X-ray beam and the detector around the instrument axis, we used a single crystal of orthoenstatite (($Mg_{1.93}Fe_{0.06}$)($Si_{1.93}$, $Al_{0.06}$)$O_6$, *Pbca* space group, *a* = 8.8117(2), *b* = 5.18320(10), and *c* = 18.2391(3) Å). Powder diffraction measurements were performed without sample rotation (still images). DIOPTAS software was used to integrate diffraction images to powder patterns.[1] Le-Bail fits of the diffraction patterns were performed with the Jana2006 software[2]. The structure was solved with the ShelXT structure solution program and refined with the Olex2 program [3,4]. The Cambridge Structural Database contains the supplementary crystallographic data for this work. These data can be obtained free of charge from FIZ Karlsruhe via www.ccdc.cam.ac.uk/structures. Details of XRD data collection and structure refinement are given in Table 1 and in the supplementary *.cif* file.

### III. RESULTS

**A. Ambient-pressure stability**

In the experiment intended to assess the stability of $CO_2$-V at close-to-ambient conditions, the sample was synthesized after initial compression to 44 GPa, where orthorhombic $CO_2$-III remained metastable [32] characterized by broad diffraction peaks (Fig. 1). After laser heating above 2300 K, new narrow peaks appeared (Fig. 1). To yield more high-pressure phase, the sample was heated over approximately a 30 x 30 μm$^2$ area. Since the Au coupler did not heat uniformly, the laser power needed to heat up the sample beyond the transformation onset varied depending on the heating point. In one area, laser heating resulted in an abrupt temperature increase in response to a regular increase in the laser power. In other areas, the temperature increased smoothly with the laser power, which was lower than in the first case. XRD patterns in the smoothly heated areas revealed the presence of $I\bar{4}2d$ $CO_2$-V, while no sign of extra peaks corresponding to other phases has been found (Fig. 1). However, one can see that in the heated area with an abrupt temperature increase there were additional peaks. XRD maps of the sample chamber show that the additional phase present in a section of the heated spot occurred at the edge of the gasket (Fig. S1 of Supplemental Material [33]). This suggests a possibility that Re compounds could be formed. It is important to note that depending on the observation spot, the ratio of intensities of the additional peaks to those of $CO_2$-V varied, and there were areas where only pure $CO_2$-V was found. A good correspondence of the $CO_2$-V peaks in all areas demonstrates that the additional peaks, which are different from those of $CO_2$-V, correspond to a distinct phase. We find that these peaks persist in the condition of low pressure and temperature. At these close-to-ambient P-T conditions, this phase can be uniquely identified as *Pbcn* $ReO_2$ [34], suggesting that this material was formed due to chemical reaction of Re gasket and $CO_2$ during laser heating (see also Ref. [35]).

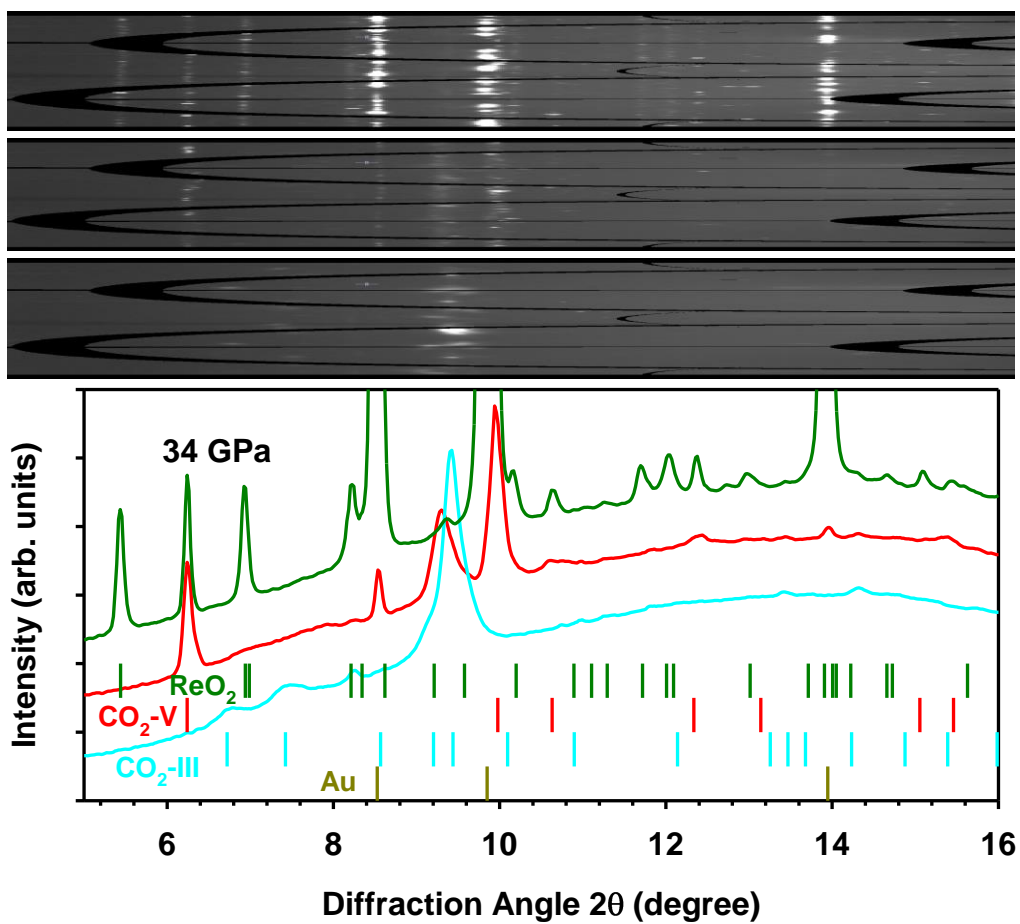

**Fig. 1.** X-ray diffraction patterns of $CO_2$ after laser heating at 44 GPa. After heating, pressure dropped to 34 GPa. Initial $CO_2$-III can be seen in unheated regions, while the patterns in the heated areas show the presence of pure $CO_2$-V (red curve) and a mixture of $CO_2$-V and *Pbcn* $ReO_2$ (green curve). Ticks correspond to the refined positions of the Bragg peak of the identified phases. The X-ray wavelength is 0.3344 Å.

XRD patterns exhibiting chemical contamination due to the presence of $ReO_2$ (Fig. 1) are similar to those reported in Ref. [6]. This pattern was indexed as orthorhombic, and the structure was inferred to be $P2_12_12_1$ based on analogy with $SiO_2$ tridymite structure. The XRD pattern reported in Ref. [6] at 48 GPa can be explained as the superposition of the patterns of $CO_2$-V and *Pbcn* $ReO_2$ (Fig. S2 of Supplemental Material [33]). This result casts doubt on the occurrence of $P2_12_12_1$ $CO_2$-V and strongly supports the existence of only one polymeric modification of $CO_2$-V with $I\bar{4}2d$ structure in the studied pressure range, in agreement with theoretical predictions [36-38] and in support of previous experimental works [7, 8, 39].

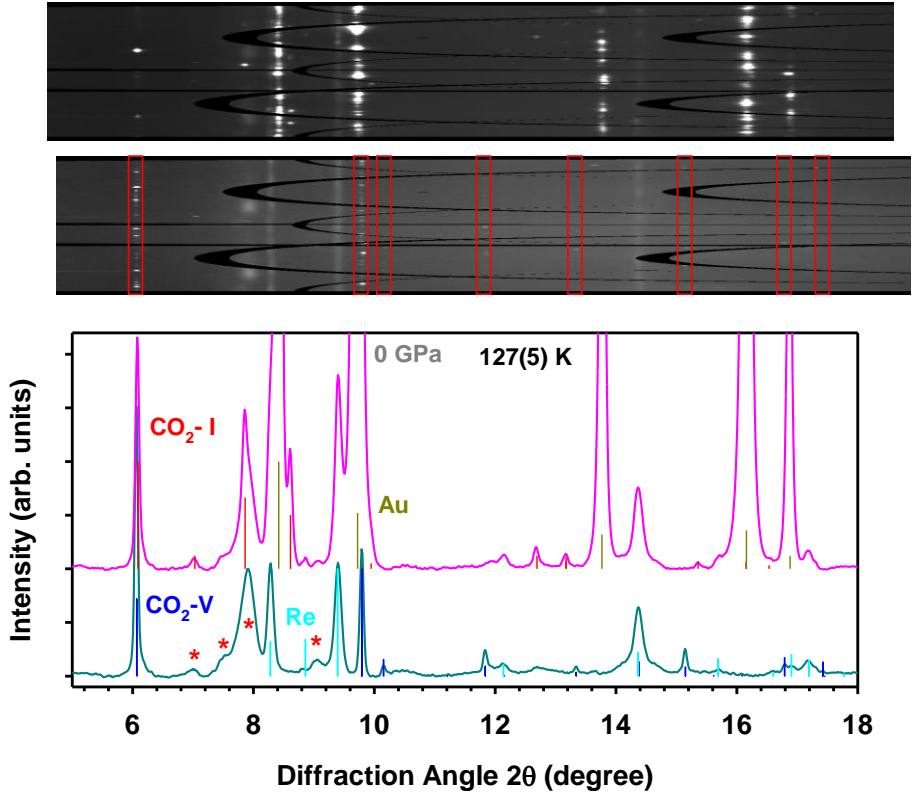

**Fig. 2.** X-ray diffraction patterns of $CO_2$ after pressure release to 0 GPa and warming up to 127(5) K. The patterns were obtained at different sample positions at nominally the same temperature just before the sublimation of $CO_2$-I (top curve) and during the final XRD mapping (bottom curve). Vertical bars correspond to the refined positions and calculated intensities of the Bragg peaks of the identified phases. Asterisks indicate reflections from the Kapton windows of the cryostat. Red rectangles box single-crystal like diffraction peaks of $CO_2$-V. The X-ray wavelength is 0.3445 Å.

Our XRD data show the stability of $CO_2$-V on cooling down to 107(9) K at 32-34 GPa and unloading down to 3 GPa along the 107 K isotherm and no other phase was recorded. On unloading to lower pressures at 107 K, we detected crystallization of $CO_2$-I. However, $CO_2$-V demonstrated (meta)stability down to 0.0(2) GPa coexisting with $CO_2$-I transforming from $CO_2$-III (Fig. 2), which remained in the high-pressure cavity after laser heating (Fig. S1 of Supplemental Material [33]). $Pbcn$ $ReO_2$ was also clearly detectable throughout the pressure and temperature range studied. After warming up to 132(5) K at close-to-ambient pressure, $CO_2$ started to sublimate, which resulted in the escape of Au heat absorber from the cavity. At this time the DAC has been totally opened, so the sample pressure would have been about 1 Pa, the pressure in the cryostat vacuum chamber. However, measurements of XRD map at 132(5) K revealed that $CO_2$-V (Fig. 2) and $Pbcn$ $ReO_2$ were still in the cavity (Fig. S1 of Supplemental Material [33]), while $CO_2$-I is only barely detectable. Please note that the quoted above temperatures were estimated based on the T dependence of the lattice parameter of $CO_2$-I at 1 Pa [40]. This was necessary because the thermocouple, which measured T directly, was positioned on the cold finger of the cryostat away from the sample position. The results presented here show the metastability of $CO_2$-V at very low pressures and temperatures below the sublimation line of $CO_2$-I.

**B. Phase stability to 286 GPa and 2500 K**

In the experiment at ultrahigh pressures, $CO_2$ was gas loaded in toroidal anvils with a 40 μm central culet with a small piece of Au, which served as a laser absorber and pressure sensor. Polymeric $CO_2$-V was synthesized at 140 GPa and above 2500 K. XRD and Raman spectroscopy measurements were performed concomitantly. At each pressure up to 286 GPa, the sample was laser heated up to at least 2500 K to release inhomogeneous stresses and to facilitate potential phase transformations. XRD was measured before and after laser heating at each pressure point except the maximum pressure, where one of the anvils failed during laser heating after initial room-temperature measurements. The sample was marginally suitable for single-crystal XRD measurements because the grains were too small and there were not sufficient observable classes of Bragg reflections to analyze the data using multigrain crystallography methods. Powder diffraction data provide a reliable identification of $I\bar{4}2d$ $CO_2$-V (Fig. 3). No change in XRD

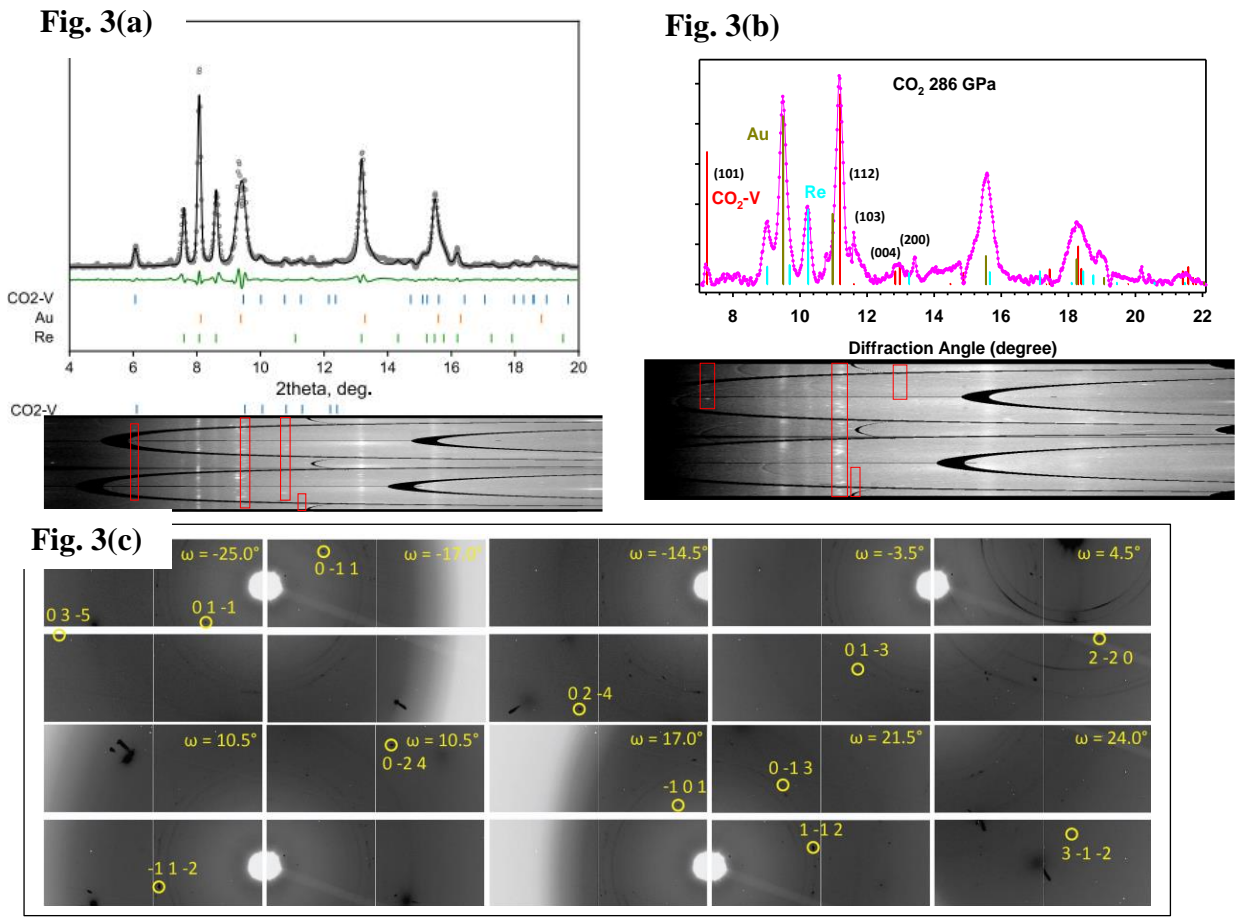

**Fig. 3.** X-ray diffraction patterns of $CO_2$ at 159 GPa (a) and 286 GPa (b,c) after laser heating. Circles are the data, and the line is the le Bail fit. Ticks correspond to Bragg peaks of the fitted

phases. The bottom panels in 3(a) and 3(b) are 2D images in rectangular coordinates. Red rectangles box single-crystal like diffraction peaks of $CO_2$-V. The panel (c) is a mosaic where the patterns are taken at different DAC rotation angles around a vertical ω-axis, demonstrating a variety of (h k l) reflections. The single-crystal reflections are circled. The X-ray wavelength is 0.2952 Å in (a) and 0.3344 Å in (b,c).

patterns were detected up to the highest pressure of 286 GPa. At this pressure, we were able to index and integrate single-crystal XRD data manually (Fig. 3(c)) further supporting the structural determination. At 286 GPa, heating up to 2500 K did not reveal any change in symmetry (Fig. S3 of Supplemental Material [33]).

Concomitant Raman experiments (Fig. 4) support the stability of $I\bar{4}2d$ $CO_2$-V up to 286 GPa as we did not detect any significant change under pressure (cf. Ref. [16]). At 159 GPa, there are 3 bands corresponding to the E, $B_2$, and $A_1$ modes in good agreement with the results of Ref. [41]. Raman spectra measured to the highest pressure exhibit the main peak, which corresponds to the $A_1$ symmetric oxygen stretching vibrations in $CO_2$ tetrahedra. Other modes, which are at least 1 order of magnitude weaker are barely seen because of a strong diamond anvil fluorescence, which increased substantially at the highest pressures. Thus, we used a 660 nm laser line to excite the spectra at 268 and 286 GPa.

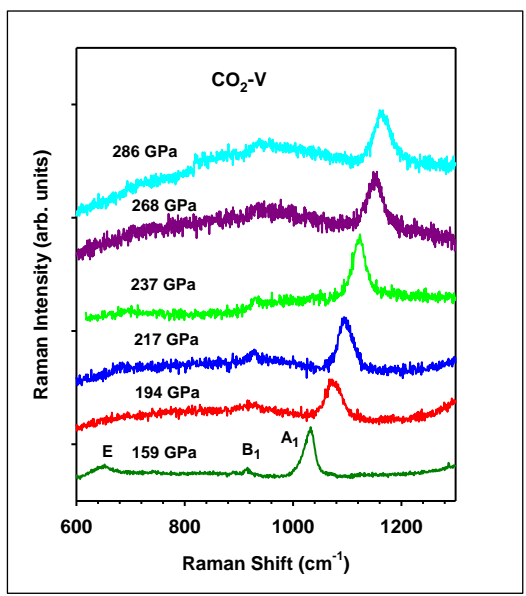

**Fig. 4.** Raman spectra at pressures between 159-286 GPa measured concomitantly with XRD patterns. The excitation laser wavelength is 532 nm for 159-237 GPa and 660 nm for 268 and 286 GPa.

The pressure dependencies of the unit cell volume (Fig. 5) and Raman frequency of the $A_1$ mode (Fig. 6) vary smoothly. Room temperature data from the present study agree well with previous investigations to lower pressures [7, 13, 39, 41]. This includes a good correspondence in the pressure behavior of the lattice parameters (Fig. S4 of Supplemental Material [33]). This agreement makes it possible to present the whole data set by a single curve as shown in Figs. 5, 6. The pressure-volume data at room temperature are best represented by a Birch-Murnaghan EOS

$$P = 1.5B_0[(V_0/V)^{7/3} - (V_0/V)^{5/3}][1 + 0.75(B_0' - 4)[(V_0/V)^{2/3} - 1]]$$

Where $B_0 = 130$ GPa and $B_0' = 4.8$ are the bulk modulus and the derivative of the bulk modulus with respect to pressure, and $V_0 = 22.8$ Å$^3$ is the volume at zero pressure. The bulk modulus (130 GPa) determined using the data of Ref. [7] is much smaller than that previously reported (365 GPa in Ref. [6]). This result is indirectly supported by our low-temperature unit-cell volume data (Fig. S4 of Supplemental Material [33]), where the bulk modulus has been found to be $B_0 = 161(4)$ GPa. Our structural data to high pressures support previous notions about an anomalous behavior of the $c$ lattice parameters [17, 41], which shows a maximum at 100-150 GPa. It is interesting that our low-temperature data (Fig. S4 of Supplemental Material [33]) also show a maximum in the $c$ vs P dependence near 15 GPa. This results in an anomaly in the unit-cell volume vs P dependence, which shows a bulge at the same pressure (Fig. S5 of Supplemental Material [33]).

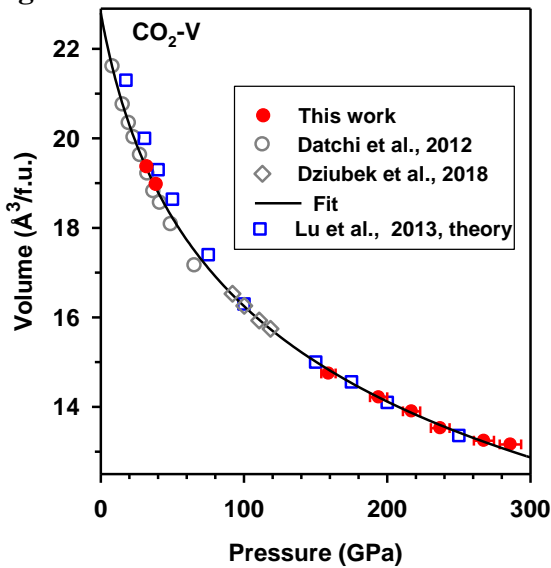
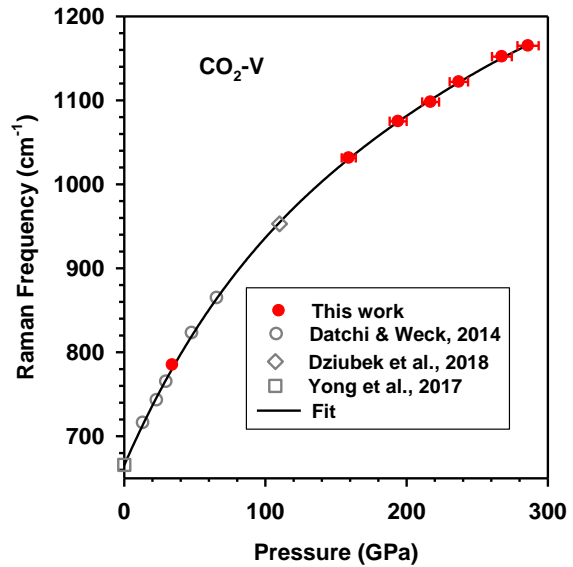

**Fig. 5.** Pressure dependence of the formula unit volume of $CO_2$-V at room temperature. Filled symbols: this work (if not shown, the error bars are smaller than experimental uncertainties), open symbols – previous works [7, 18, 41]. Solid black curve representing the experimental data in the whole pressure range is a Birch-Murnaghan EOS with the following parameters: $V_0$=23.8 Å$^3$, $B_0$=130 GPa, $B_0'$=4.8.

**Fig. 6.** Pressure dependence of the $A_1$ Raman mode frequency of $CO_2$-V at room temperature. Filled symbols: this work (if not shown, the error bars are smaller than experimental uncertainties), open symbols – previous works. Solid black curve representing the experimental data in the whole pressure range is a fourth-order polynomial.

The frequency versus pressure data is well represented by a fourth order polynomial with $a_0$=665.5 cm$^{-1}$, $a_1$=3.829 cm$^{-1}$ GPa$^{-1}$, $a_2$=-0.0144 cm$^{-1}$ GPa$^{-2}$, $a_3$=3.632*10$^{-5}$ cm$^{-1}$ GPa$^{-3}$, $a_4$=-3.941*10$^{-8}$ cm$^{-1}$ GPa$^{-4}$. These data allow us to directly determine the volume dependent mode Grüneisen parameter of the $A_1$ mode $\gamma_{A1} = -\partial \ln \omega_{A1}/\partial \ln V$, which we find to be pressure independent within the accuracy of the available data and equal to 1.0(1). This result establishes a "normal" behavior of covalent C-O bonds (similar to diamond [42]) in $CO_4$ tetrahedra in $CO_2$-V (e.g., Ref. [43]) supporting its stability under high pressure.

**C. Chemical stability**

In the experiment on the CO-$N_2$ system, mixed CO and $N_2$ gases were loaded in a DAC with an Au coupler for laser heating. Pressure was initially monitored by observing a change in the sample appearance [24]. To avoid photochemical transformations [44, 45], no optical laser-assisted measurements were performed below 20 GPa; pressure was controlled via Raman of the stressed diamond anvil at higher pressures. The sample was heated at 39 GPa up to 2500 K via coupling of the laser radiation to Au pieces in the high-pressure cavity. The products of chemical/physical transformations were studied by means of single-crystal and powder X-ray diffraction at beamline 16 ID-B of the Advanced Photon Source.

Before laser heating, there were no Bragg peaks from the sample in accordance with Ref. [24] because mixed CO and $N_2$ solids are amorphous at these conditions. This has been verified by making XRD maps of the sample chamber. After laser heating, Bragg peaks were found in the heated areas which were identified as $CO_2$-V and diamond (Fig. 7). Lattice parameters and unit cell volumes are in good agreement with our data and those reported previously [17, 20, 41] (Fig. 5, Fig. S4 of Supplemental Material [33]).

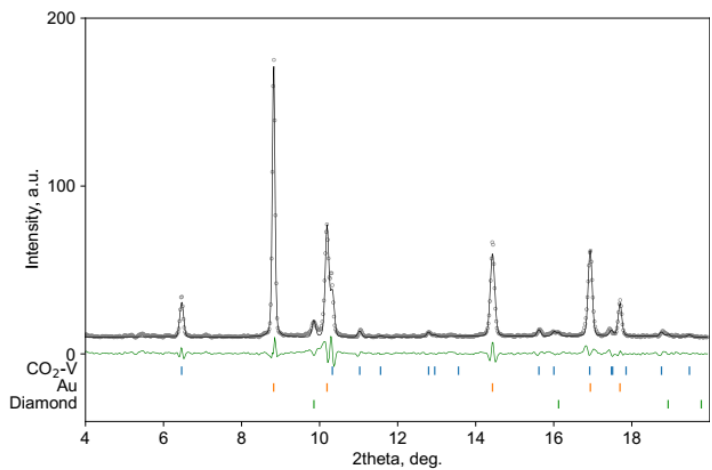

**Fig. 7.** X-ray diffraction patterns of $CO_2$ synthesized at 38.6 GPa after laser heating of CO-$N_2$ mixture at 39 GPa. Circles are the data, and a line is the le Bail fit. Ticks correspond to Bragg peaks of the fitted phases. The X-ray wavelength 0.3445 Å.

This phase assignment has been uniquely confirmed by single-crystal diffraction measurements (Fig. 8, Table S1 of Supplemental Material [33]). Laser heating promotes the decomposition of CO to $CO_2$-V and diamond:

$$2CO \rightarrow CO_2 + C$$

The $CO_2$-V phase was detected based on powder and single-crystal X-ray diffraction analysis, while diamond peaks were found on the powder patterns only. Very little if any crystalline $\varepsilon$-$N_2$ diffraction signal was observed in our XRD pattern in contrast to Ref. [24]. The Bragg peaks assigned to a new polymeric $P4_3$ $CON_2$ solid in this work can be explained by the presence of $CO_2$-V (Fig. S6 of Supplemental Material [33]) in their laser heated samples.

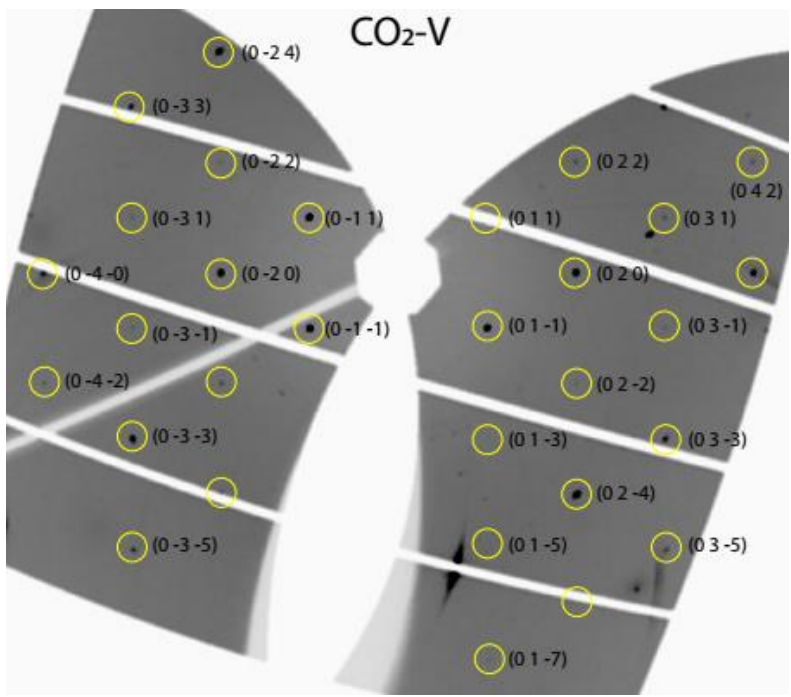

**Fig. 8.** Reconstructed reciprocal lattice plane (0*kl*) of $CO_2$-V at 38.6 GPa.

## IV. DISCUSSION

Our combined XRD-Raman investigations over a wide P-T range shed light on the physics and chemistry of C-O compounds at extreme conditions. Early considerations about the stability of various atomic and electronic configurations were driven by the assumed similarity of $CO_2$ with isoelectronic compounds $SiO_2$ and $GeO_2$ [6, 22]. In particular, compounds with octahedral carbon coordination were expected to become stable in the pressure range within the reach of DAC techniques, as $SiO_2$ and $GeO_2$ are known to demonstrate transformations into octahedrally-coordinated structures at very moderate pressure conditions. Moreover, such octahedral modification (e.g., stishovite) is metastable at ambient conditions, while recovery of tetrahedrally coordinated carbon in polymeric $CO_2$ has remained controversial.

Yong et al. [13] reported molecular dynamics (MD) simulations on $CO_2$-V at 300 K while reducing pressure which show the absence of a back transformation into molecular $CO_2$ solids for long simulation times. In addition, the phonon structure calculations of tetrahedrally coordinated $CO_2$ at ambient pressure showed the dynamical stability of the structure, thus suggesting that it can be recovered. However, the recovery may not be possible because of kinetic reasons. For the experiments of Ref. [13], the temperature at which recovery was attempted was likely not sufficiently low, so the back transformation into a molecular phase could occur. The experiments of Ref. [13] appear to have a large pressure uncertainty [20], thus making unreliable the initial claim about the recovery at 191 K. Large hysteresis loops are common for transformations of molecular crystals into polymeric states (e.g., Ref. [46]), so it is not surprising if the back transformation would occur close to ambient pressure. Moreover, it is not expected that the recovered polymeric phase would be stable above the melting line (or sublimation line in case of $CO_2$) for the molecular crystal [47]. In the case of $CO_2$, recovery of $CO_2$-V below 217 K is expected if $CO_2$-V sublimates along a similar curve to $CO_2$-I at pressures below 5.2 atm (0.52 MPa) (Fig. 9). At 191 K, where the recovery was attempted in Ref. [13], the lower bound of pressure of stability of the solid is close to 1 atm. Thus, we suggest that even though it has not been definitively proven in previous studies [16], $CO_2$-V recovery at 191 K of Ref. [13] to atmospheric pressure still remains a possibility.

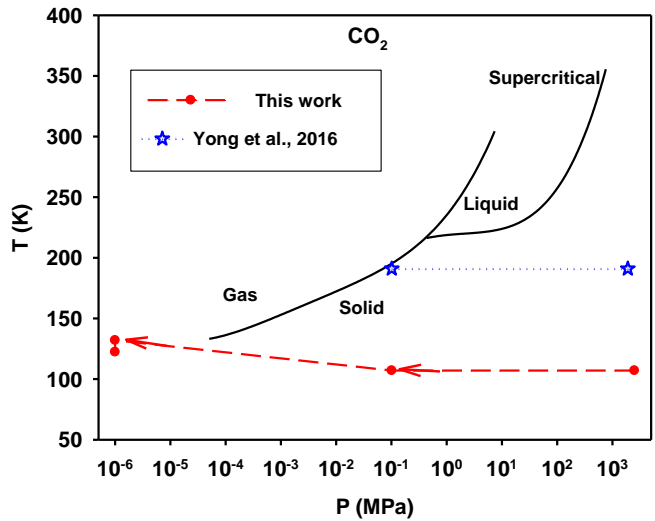

**Fig. 9.** Phase diagram of $CO_2$ at low pressures. Symbols show the estimated P-T conditions of observations for $CO_2$-V in this work and Ref. [13]. The arrows show the P-T path of this work.

In contrast, our experiments explored a lower-temperature decompression pathway at 107(9) K. Only $CO_2$-V has been detected down to 3 GPa in the sample regions where it was synthesized, while there were only weak peaks of $CO_2$-III in the untransformed regions. While $CO_2$-III transforms to $CO_2$-I at 3 GPa, $CO_2$-V appears to remain untransformed (Fig. S1 of Supplemental Material [33]). This situation holds down to the lowest pressure of 0.1(1) GPa which was recorded using Au for a pressure sensor. $CO_2$-V remained stable at these conditions based on our XRD measurements (Fig. S1 of Supplemental Material [33]). Warming up to 132(5) K in the unsealed sample chamber resulted in partial $CO_2$ sublimation. At 122-132 K, a mixture of $CO_2$-V and $CO_2$-I phases was observed at vacuum pressure in the cryostat (<1 Pa) (Fig. S1 of Supplemental Material [33]). Thus, our experiments clearly demonstrate that polymeric $CO_2$-V is quenchable to ambient pressure- arguably the only example for a simple molecular system.

Several recent experimental and theoretical investigations explored the stability of molecular and polymeric phases of $CO_2$ [17, 18, 41, 48, 49]. These investigations were addressing reports about molecular breakdown [2], molecular disproportionation and ionization [16], and amorphization of $CO_2$-V [16] in the limits of high P and high T. While below 120 GPa these phenomena were extensively investigated experimentally [17, 41] demonstrating exceptional stability of $CO_2$-V at 40-120 GPa up to 6000 K, the higher-pressure behavior remained less explored. Here we extended the probed pressure range up to 286 GPa demonstrating the stability of $CO_2$-V up to these conditions and high temperatures up to at least 2500 K. These results show that the stability field of $CO_2$-V [17, 41] holds to much higher pressures. No amorphization is observed above 220 GPa at 300 K (Figs. 3-6) as reported in Ref. [16]. Instead, we find normal compression behavior with no sign of instability. The theoretically predicted transformation into another tetrahedrally coordinated $CO_2$ phase above 200 GPa [19] and 285 GPa [18] was not detected likely because the experiment did not reach the transition pressure, which is higher than predicted. The transition to 6-coordinated carbon structures could not be reached here because it is expected at much higher pressures near 1 TPa [18]. The reason why such high pressures are required compared to heavier isoelectronic analogs $SiO_2$ and $GeO_2$ is likely due to a difference in electronic structure of C versus Si and Ge, where the C core does not possess *p*-electrons, making this atom much more compact. This allows C to occupy interstitial sites in the closed packed O sublattice unlike in $SiO_2$ [19], making this structure very stable over a wide pressure range.

It is important to note that, compared to Refs. [17, 41], our experiments employed a different laser heating method, where a near IR laser was coupled to an Au heat absorber. As the results agree well concerning the structure, chemical stability and EOS, we conjecture that other laser absorbing materials such as Ir [2], Pt [16], $SiO_2$, ruby, and Re gasket [6, 50, 51] could result in unwanted chemical reactions including decomposition into diamond and molecular $O_2$ [2, 51]. Here we documented the formation of *Pbcn* $ReO_2$ (Fig. 1 and Fig. S2 of Supplemental Material [33]), likely adding difficulty in experiments [6, 13], and subsequent misinterpretation of the structure of polymeric $CO_2$. Partial decomposition of $CO_2$ into $O_2$ and carbon in the presence of Ir [2] was likely a catalytic effect as it was observed even at lower T in the presence of well-known catalyst Pt [35]. Because the X-ray and Raman cross sections of Re compounds and ε-$O_2$ at 40-60 GPa, respectively, are relatively high compared to $CO_2$, even small amounts of the reaction products may be sufficient to produce a strong signal confusing the reported results. Other, yet unidentified contaminants (e.g., nitrogen) may be responsible for ionized $CO_2$ at 85 GPa [16].

Finally, we discuss the possible chemical reactivity of the C-O subsystem in the presence of other elements with the example of N. Thermodynamically stable ternary compounds in this system are rare because the end member and binary compounds are very stable [23, 25]. Theoretical calculations in the system CO-$N_2$ [52, 53] predicted very few stable polymeric compounds above 35 GPa. Out of them the $P4_3$ polymeric structure is selected as the most interesting because it forms a 3D framework lattice with four coordinated C, three coordinated N, and two coordinated O. Synthesis of this compound has been reported in Ref. [24]. However, the powder XRD pattern of Ref. [24] can be better explained by a superposition of $CO_2$-V and ε-$N_2$ (Fig. S6 of Supplemental Material [33]) casting doubt on this interpretation. Our experiments clearly demonstrate synthesis of $CO_2$-V confirmed by powder (Fig. 7) and single-crystal XRD measurements (Fig. 8, Table S1 of Supplemental Material [33]). These observations and signs of diamond formation (Fig. 7) show that high-pressure reactions in the CO-$N_2$ system in a laser heated DAC follow the pathway of CO decomposition. This complicates the task of CO-$N_2$ compound synthesis in a DAC.

## V. CONCLUSIONS

Our experiments demonstrated the exceptional stability of the tetrahedrally coordinated polymeric compound $CO_2$-V over a pressure and temperature range of 0-286 GPa and 100-3000 K, respectively. The theoretically predicted transformation of $CO_2$ at 200-285 GPa into a layered polymeric $P4_2/nmc$ $CO_2$ structure has not been detected and is thus expected to occur at higher pressures. Polymeric $CO_2$-V can be recovered at ambient pressure at temperatures below 122 K. Stoichiometries other than $CO_2$ (for example, CO) tend to chemically react forming $CO_2$-V above 35 GPa. Thus, complex synthetic routes are needed to make C-O bearing polynitrides as high energy density and ultrahard materials.

# Supplemental Material

# Assessing the stability fields of molecular and polymeric $CO_2$

Alexander F. Goncharov[1], Elena Bykova[1,2], Maxim Bykov[1,3], Eric Edmund[1], Jesse S. Smith[4], Stella Chariton[5], and Vitali B. Prakapenka[5]

[1] Earth and Planets Laboratory, Carnegie Institution for Science, Washington, DC 20015, USA

[2] Bayerisches Geoinstitut, University of Bayreuth, Universitätsstrasse 30, D-95447, Bayreuth, Germany

[3] Institute of Inorganic Chemistry, University of Cologne, Greinstrasse 6, 50939 Cologne, Germany

[4] HPCAT, X-ray Science Division, Argonne National Laboratory, Argonne, Illinois 60439, USA

[5] Center for Advanced Radiation Sources, The University of Chicago, Chicago, Illinois 60637, USA

This pdf file contains 6 Supplemental Figures and 1 Supplemental Table.

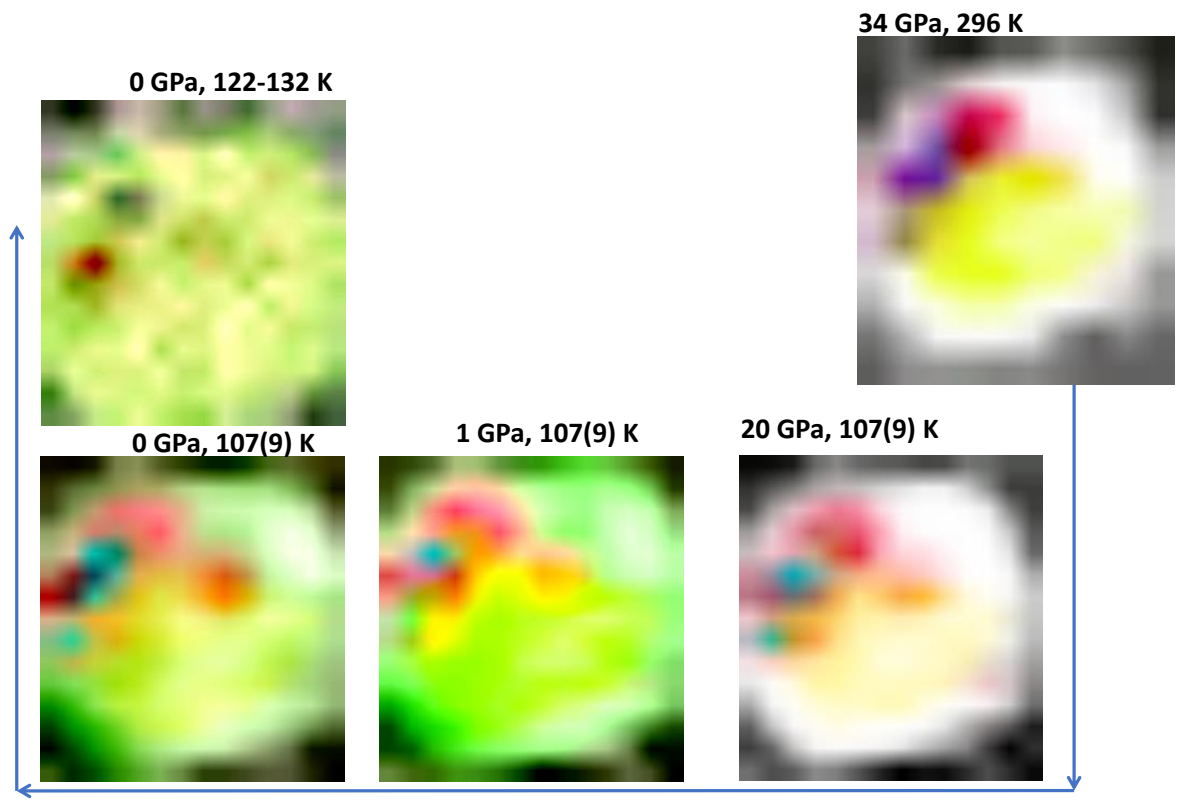

**Fig. S1**. X-ray map of reaction products after laser heating of $CO_2$ in a DAC cavity with Au heat absorber on the P-T pathway (shown by the arrows) exploring the recovery $CO_2$-V to ambient conditions. Red colors correspond to $CO_2$-V, blue colors- $ReO_2$, yellow- Au, green – $CO_2$-I.

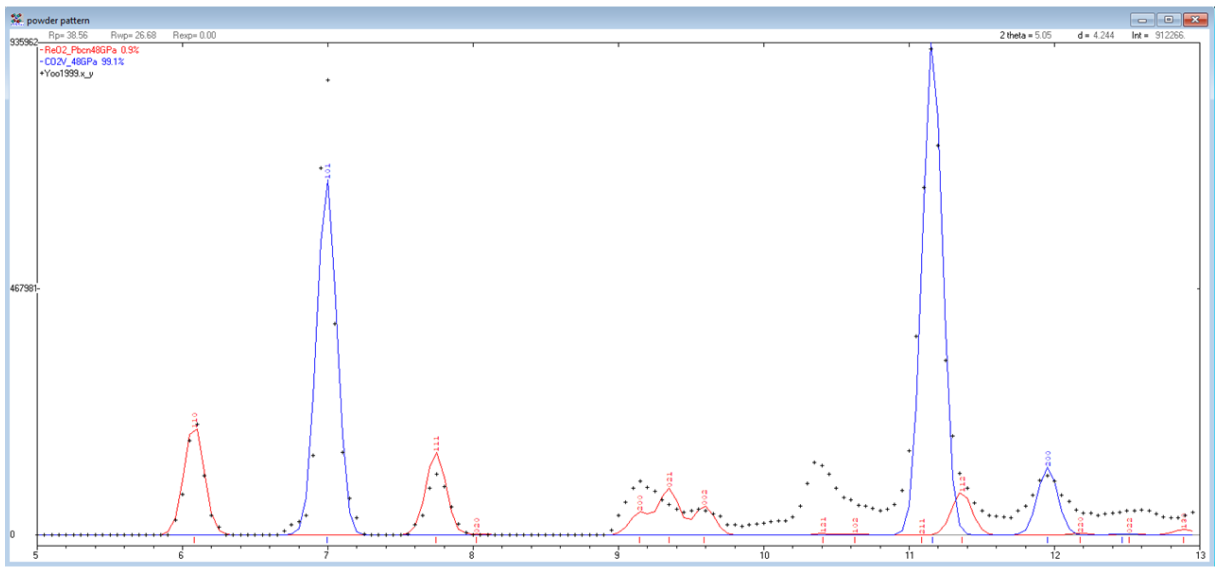

**Fig. S2.** Powder diffraction pattern of polymeric $CO_2$ at 48 GPa reported by Yoo et al. [1] (dots) superimposed with the calculated XRD [2] of $CO_2$-V (a=3.59 Å, c=5.885 Å) and *Pbcn* $ReO_2$ [3] (a=4.688 Å, b=5.345 Å, c=4.472 Å) shown as blue and red curves, respectively. The X-ray wavelength is 0.3738 Å.

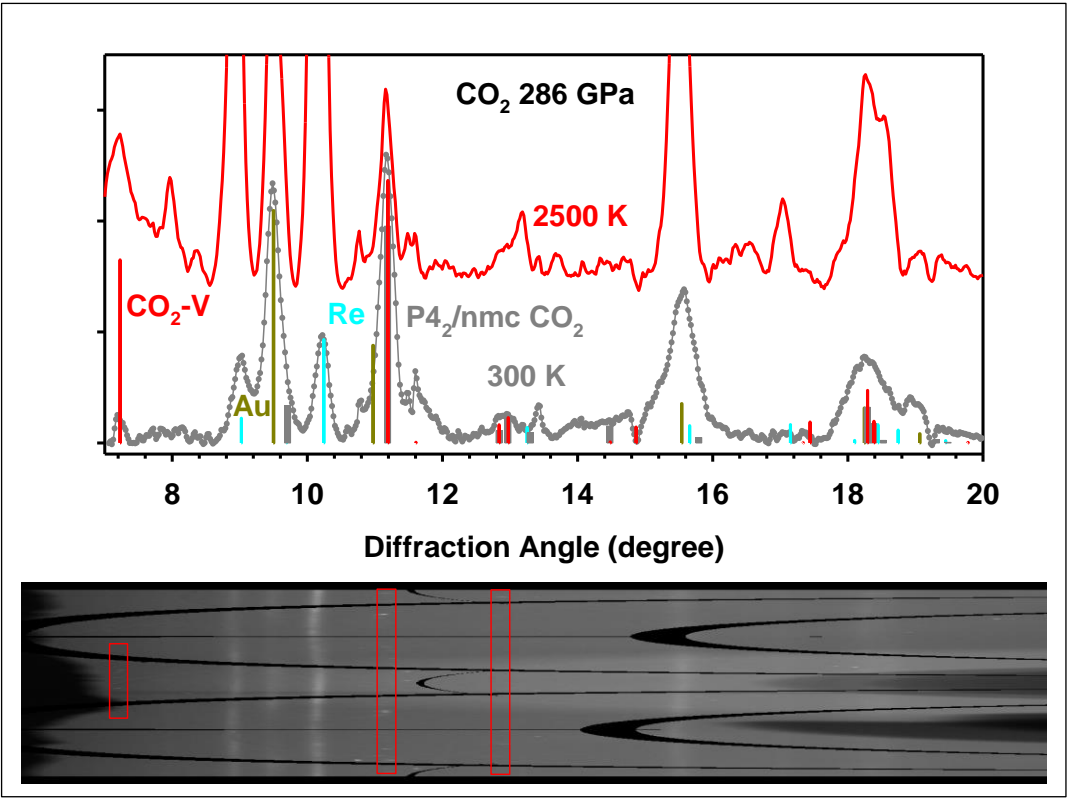

**Fig. S3.** X-ray diffraction patterns of $CO_2$ at 286 GPa at 300 K and laser heated to 2500 K. Ticks correspond to Bragg peaks of the fitted phases. Gray bars correspond to a theoretically predicted $P4_2/nmc$ phase with the same density as that of $CO_2$-V. The bottom panel is a 2D image in rectangular coordinates at 2500 K. Red squares in box single-crystal like diffraction peaks of $CO_2$-V. The X-ray wavelength is 0.3344 Å.

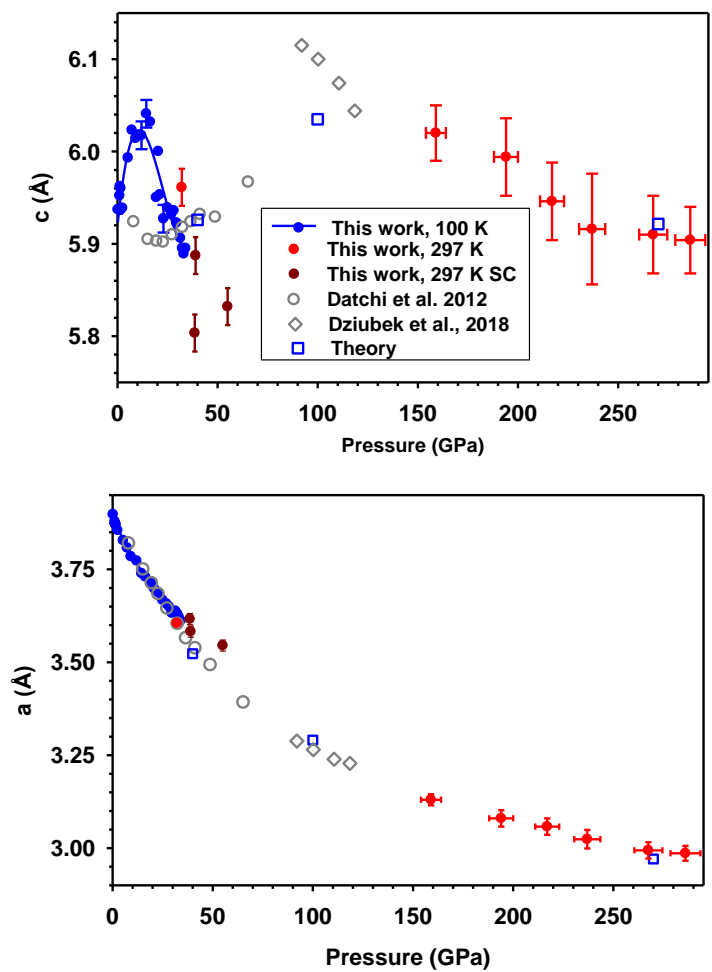

**Fig. S4.** Pressure dependence of the lattice parameters of $CO_2$-V at room and low temperatures. Filled symbols: this work (if not shown, the error bars are smaller than experimental uncertainties), open symbols – previous works [4-6]. Open blue squares are theory data from Refs. [6-8].

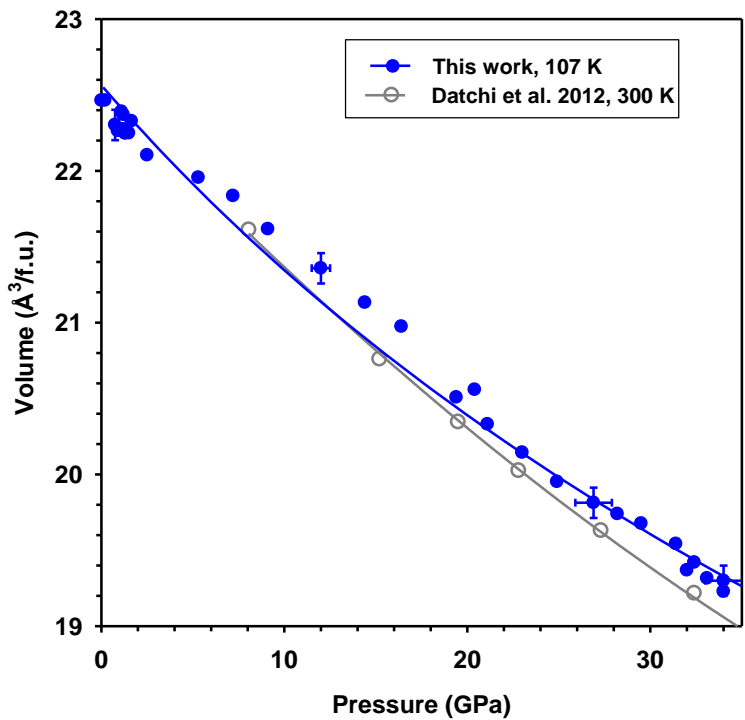

**Fig. S5.** Pressure dependence of the formular unit volume of $CO_2$-V at low temperature. Filled symbols: this work, open symbols – previous work [4]. Solid black curve representing the experimental data in the whole pressure range is a Birch-Murnaghan EOS with the following parameters: $V_0$=22.56(4) Å$^3$, $B_0$=161(4) GPa, $B_0'$=4 (fixed).

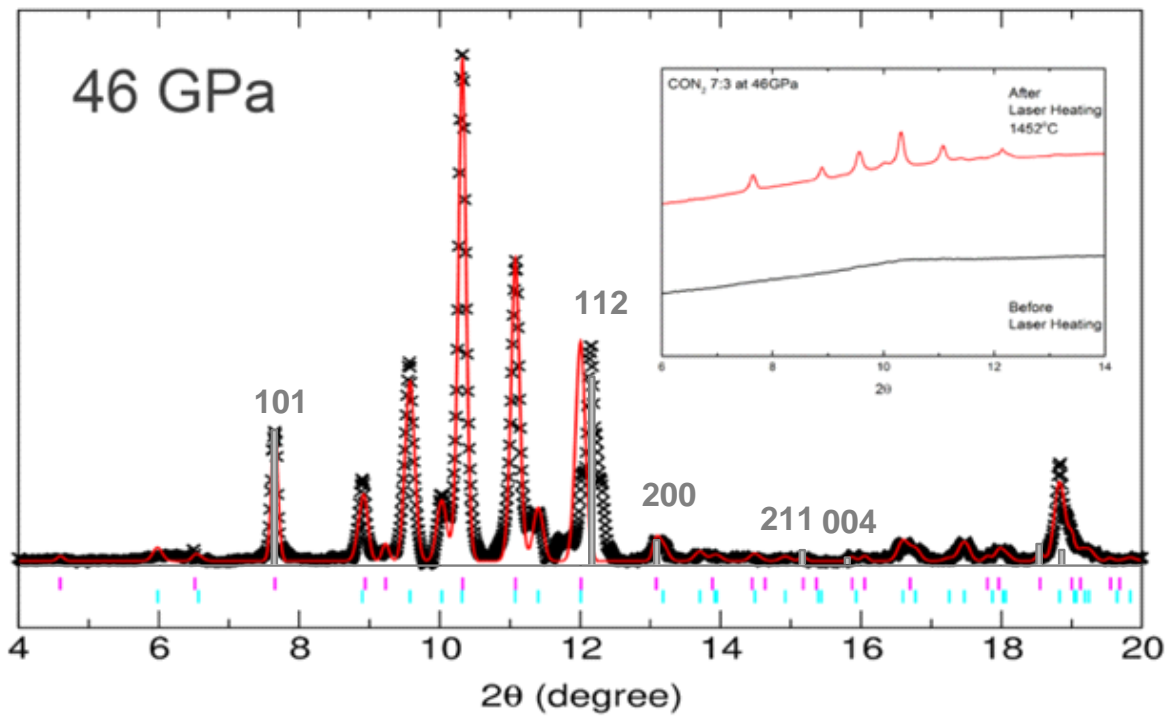

**Fig. S6.** Powder diffraction pattern laser heated CO-N$_2$ sample at 46 GPa reported by Yoo et al. [9]. Red curve is the proposed in Ref. [9] Rietveld refinement of $P4_3$ CON$_2$ solid; red and blue ticks correspond to $P4_3$ CON$_2$ solid and ε-N$_2$, respectively. The data (crosses) can be better explained by mixed ε-N$_2$ (blue ticks) and CO$_2$-V solids (gray bars, the strongest peaks are labeled). The X-ray wavelength is 0.4062 Å.

**Table S1. Crystal structure determination of $CO_2$-V at 39 GPa following chemical reaction in CO-$N_2$ system initiated by laser heating at 39 GPa.**

| Crystal data | |
|---|---|
| Chemical formula | $CO_2$ |
| $M_r$ | 44.01 |
| Crystal system, space group | Tetragonal, $I\bar{4}2d$ |
| Temperature (K) | 295 |
| $a$, $c$ (Å) | 3.6162 (5), 5.8034 (15) |
| $V$ (Å$^3$) | 75.89 (3) |
| $Z$ | 4 |
| Radiation type | Synchrotron, $\lambda$ = 0.34453 Å |
| $\mu$ (mm$^{-1}$) | 0.11 |
| Crystal size (mm) | 0.002 × 0.002 × 0.002 |
| **Data collection** | |
| Diffractometer | Customized ω-circle diffractometer |
| Absorption correction | Multi-scan *CrysAlis PRO* 1.171.41.122a (Rigaku Oxford Diffraction, 2021) Empirical absorption correction using spherical harmonics, implemented in SCALE3 ABSPACK scaling algorithm. |
| $T_{min}$, $T_{max}$ | 0.423, 1.000 |
| No. of measured, independent and observed [$I > 2\sigma(I)$] reflections | 61, 38, 36 |
| $R_{int}$ | 0.085 |
| $(\sin \theta/\lambda)_{max}$ (Å$^{-1}$) | 0.715 |
| **Refinement** | |
| $R[F^2 > 2\sigma(F^2)]$, $wR(F^2)$, $S$ | 0.085, 0.239, 1.20 |
| No. of reflections | 38 |
| No. of parameters | 8 |
| $\Delta\rho_{max}$, $\Delta\rho_{min}$ (e Å$^{-3}$) | 0.69, −0.79 |